\documentclass[aps,preprint,preprintnumbers]{revtex4}
\usepackage{graphicx}
\usepackage{amsmath,amssymb}
\usepackage{bm}
\begin{document}
\title{Low-energy Landau levels of AB-stacked zigzag graphene ribbons}
\author{Y. C. Huang}
\affiliation{\footnotesize Center for General Education, Kao Yuan University, 821 Kaohsiung, Taiwan}
\author{C. P. Chang}
\email{t00252@mail.tut.edu.tw}
\affiliation{\footnotesize  Center for General Education, Tainan  University of  Technology, 710 Tainan, Taiwan}
\author{W. S. Su}
\affiliation{\footnotesize  Center for General Education, Tainan  University of  Technology, 710 Tainan, Taiwan}
\author{M. F. Lin}
\email{mflin@mail.ncku.edu.tw}
\affiliation{\footnotesize Department of Physics, National Cheng Kung University, 701 Tainan,Taiwan}
\begin{abstract}
Low-energy Landau levels of AB-stacked zigzag graphene ribbons in the presence of a uniform perpendicular magnetic field (\textbf{B}) are investigated by the Peierls coupling tight-binding model. State energies and associated wave functions are dominated by the \textbf{B}-field strength and the $k_z$-dependent interribbon interactions. The occupied valence bands are asymmetric to the unoccupied conduction bands about the Fermi level.  Many doubly degenerate Landau levels and singlet curving magnetobands exist along $k_x$ and $k_z$ directions, respectively. Such features are directly reflected in density of states, which exhibits a lot of asymmetric prominent peaks because of 1D curving bands. The $k_z$-dependent interribbon interactions dramatically modify the magnetobands, such as the lift of double degeneracy, the change of state energies, and the production of two groups of curving magnetobands. They also change the characteristics of the wave functions and cause the redistribution of the charge carrier density. The $k_z$-dependent wave functions are further used to predict
the selection rule of the optical transition.
\end{abstract}
\maketitle

\section {Introduction}
Carbon-based materials, such as, graphite, carbon nanotubes,\cite{Iijim}  two-dimensional (2D) graphene and few-layer graphenes,\cite{Novoselov01,Bunch,Zhang01,Wu} and graphene ribbons, have been widely studied both experimentally and theoretically. Graphene ribbons could be synthesized by using heat treatment,\cite{Murakami,Prasad01,Zhang02} pulsed-laser deposition technique,\cite{Yudasaka}  patterning epitaxially grown graphenes,\cite{Berger01,Berger02}   tailoring exfoliated graphenes\cite{Novoselov02,Zhang03} by scanning tunneling microscopy,\cite{Hiura} or chemical vapor deposition.\cite{Jessica}  A one-dimensional (1D) graphene ribbon  is obtained by cutting a 2D graphene, planar hexagonal lattices of carbon atom,  along the longitudinal direction. As a result of the special geometric structure, graphene ribbons have motivated many interesting studies on magnetic properties,\cite{Wakabayashi,Prasad02,Ritter,
Huang01,Huang02}  optical properties, \cite{Huang01,Huang02,Lin01} electronic excitations,\cite{Lin02} or electronic properties.\cite{Chang01} The objective of this work is to investigate the  magnetoelectronic structures of Bernal graphene ribbons.

1D  zigzag and armchair graphene ribbons, sided with two parallel zigzag and armchair structures  along the longitudinal direction, respectively,  are intensively studied.  Many theoretical studies predict that a zigzag ribbon has peculiar edge  states. Such edge states produce  flatbands at low energy and  give rise to a conspicuous peak in density of states.\cite{Nakada} An armchair ribbon does not exhibit such states. Its electronic properties, such as the band gap, depend on the ribbon width.  Furthermore,  2D multilayer  zigzag (armchair) graphene ribbons are the stack of infinite identical 1D  zigzag (armchair) graphene ribbons along the stacking  direction  [figure 1].  This stacked system exhibits the  anisotropic energy dispersions between in-ribbon-pane  and the stacking direction for the intraribbon coupling is much stronger than the interribbon interaction. Moreover, the interribbon interactions also modify the in-ribbon-pane electronic properties, such as state energies, energy dispersions, band-edge states, and size of band gap.
The study results show that the geometrical structures (the stacking types (AA or AB stacking), ribbon width, and ribbon edge structure) have a significant effect
on the electronic and  optical properties of  multilayer graphene ribbons.\cite{Shyu01,Chiu,Chang02}  The optical measurement might serve a method to  determine the geometrical structures of  graphene ribbons.\cite{Ferrari}

When  a monolayer graphene ribbon is submitted to  a perpendicular magnetic field \textbf{B},  its magnetoelectronic properties, e.g., the magnetic bands  and the related wave functions,  are   determined by the competition between the  magnetic confinement and the quantum confinement.\cite{Brey,Huang01,Huang02}  When the spatial extent of Landau wave functions is smaller than ribbon width, i.e., the magnetic confinement  predominates over the quantum confinement, the Landau states exist and the state energy follows  $E \propto{\sqrt {|n| {\rm B}}}$
(n meaning the subband index).\cite{Huang02}  Such a simple characteristic is absent from a bilayer Bernal graphene ribbon. According to literature,\cite{McCann,Henriksen,Huang03,Plochocka} bilayer  graphene ribbons  exhibit  different ${\rm B}$-dependent Landau levels as a result of the interribbon interaction. In the energy region  $|E| \le$ 50 meV,  the Landau-level energies  are linearly dependent on the magnetic field strength. In the higher energy region,  they deviate from the ${\rm B}$-dependence to the $\sqrt {\rm  B}$-dependence with the increase of the field strength.  Based on the findings mentioned above, Bernal graphene ribbons (AB-stacked graphene  ribbons) are thus expected to have different magnetoelectronic properties due to the different geometrical structure and interribbon interactions.

On the other hand,  experimental measurements confirm that Landau levels of monolayer graphene and epitaxial graphene obey the relation  $E\propto{\sqrt {|n| {\rm B} }}$, Landau levels of Dirac fermions.\cite{Plochocka,Jiang,Sadowski}  Interestingly, Landau levels of Dirac fermions in graphite are experimentally observed.\cite{Li} Recently, Far infrared magnetotransmission measurements on a thin graphite show the ${\sqrt {\rm B}}$-dependent absorption lines  at the $H$ point of graphite.\cite{Orlita}  The magneto-transmission  of graphite at $K$ point, where interlayer interactions are maximum along the $HKH$ edge of Brillouin zone, exhibits  $B$-dependent Landau levels.\cite{Orlita} Both the theoretical and experimental studies on the  electronic and magneto-optical properties of graphitic systems motivate us to study the magnetoelectronic properties of Bernal graphene ribbons. The organization of the present paper is as follows. The analytic Hamiltonian matrix elements of the tight-binding method for magnetoelectronic properties of  AB-stacked graphene  ribbons  are first  derived in Sec. 2. Then, Sec. 3  investigates the effects of magnetic fields and the interlayer interactions on the band dispersions, Landau plot, and wave functions, following by the prediction of
the selection rule of the optical transition. Lastly,
conclusions are drawn in  Sec. 4.

\section {Theory}

The  AB-stacked zigzag graphene ribbon is chosen for the model study. It has hydrogen-terminated zigzag edges along the longitudinal ($\hat x$) direction [Fig. 1(a)]. The ribbon width is defined by the number ($N_y$) of zigzag lines along the $y$ axis. $B$ and $A$ atoms are different sublattices of biparticles hexagonal lattice of a graphene ribbon. There are 2$N_y$ carbon atoms in a  primitive cell of a monolayer graphene ribbon. To obtain  the 2D AB-stacked zigzag graphene ribbon, the same 1D zigzag ribbons are piled in the ABAB sequence along the $z$ axis with the stacking distance $I_z= 3.35$ {\rm \AA}. The \textit{A} atoms have the corresponding atoms on the $x-y$ plane directly above or below them, while  the projections of the $B$ atoms are located at the center of the hexagonal rings directly above or below them [Fig. 1(b)].

In the presence of a perpendicular magnetic field ${\bf B}=(0, 0, {\rm B})$, the electronic properties of  AB-stacked zigzag ribbons are modeled by the Peierls coupling tight-binding model. The Hamiltonian is
\begin{eqnarray}
{ \bf  H } =\sum_{i,j} (\gamma_{i,j}  e^{i 2 \pi \theta_{i,j} }
c_i^+ c_j + H. c. ) ,
\end{eqnarray}
where the subindices $i,j$ denote the summation  over all sites in a primitive unit cell. $c_i^+ ( c_j)$ is the creation (annihilation) operation, which generates (destroys) an electron at $i$ ($j$) site. The details of the atom-atom interactions\cite{Charlier} $\gamma_{i,j}=\gamma_0,~\gamma_1,\cdots,\gamma_6$ are described in Fig. 1(b).  $e^{i 2 \pi \theta_{i,j}}$ is the Peierls phase shift due to the applied magnetic field. $\theta_{i,j}$ is the line integral of  vector potential ${\bf A}$ from  $i$ to $j$ in a unit of the flux quantum $ \Phi_0 =  ch /e$. To keep the translation invariance along the $x$ axis, within the Landau gauge, the vector potential ${\bf A}=(-{\rm B}y, 0, 0)$ is deliberatively  adopted. In this gauge,  AB-stacked zigzag ribbons have two ribbons or 4$N_y$ carbon atoms in a  primitive cell. The first Brillouin zone is a rectangle defined by $-\pi/I_x \le k_x \le  \pi/I_x$, and $-\pi/2I_z \le k_z \le \pi/2I_z$.

The tight-binding Bloch  function is the linear combination of basis $|A^1_m \rangle,~ |B^1_m\rangle, ~|A^2_m \rangle$, and $|B^2_m \rangle$, which are the linear superposition of the  $2p_z$ orbitals located at $A^1_m,~ B^1_m, ~A^2_m$, and $B^2_m$ carbon atoms. The wave function is
\begin{eqnarray}
|\Phi(k_x,y,k_z)\rangle=\sum_{m=1}^{N_y}  { \emph{a}}_{A^1_m} |A^1_m \rangle + {\emph{b}}_{B^1_m}|B^1_m \rangle +\sum_{m=1}^{N_y}{\emph{a}}_{A^2_m} |A^2_m\rangle+  { \emph{b}}_{B^2_m} |B^2_m\rangle\text{,}
\end{eqnarray}
where wave function coefficients ${\emph{a}}_{A^1_m}$,
${\emph{b}}_{B^1_m}$,   ${\emph{a}}_{A^2_m}$  and
${\emph{b}}_{B^2_m}$ are the site amplitudes.

The Hamiltonian representation is a $ 4N_y \times  4N_y$ Hermitian matrix and could be regarded as
\begin{eqnarray}
{ \bf  H } =\left[\begin{array}{cc}
                    h_1 & h_{12}\\
                    h_{21}& h_2
                     \end{array}  \right ],
\end{eqnarray}
where each Hermitian block matrix $h_1, h_2, h_{12}$ or $h_{21}$ has $2N_y \times  2N_y$ elements.  The block matrix $h_1$, Hamiltonian representation of the lower ribbon, is a tridiagonal matrix and its matrix elements $h_1(i,j)$ are:
\begin{eqnarray}
\begin{cases}
h_1(2m-1,2m-1) = \gamma_6\,+\beta^2\gamma_5\,/2,\\
h_1(2m,2m) = \beta^2\gamma_2/2,
h_1(2m-1,2m)=h_1(2m,2m-1)=2\gamma_0\,
cos\{\sqrt 3\,bk_x/2 - \pi(m-[N])\Phi\},\\
h_1(2m,2m+1)=h_1(2m+1,2m)=\gamma_0,
\end{cases}
\label{h1}
\end{eqnarray}
where $m=1$, 2, $\cdots$, $N_y$ and $\beta = 2cos(k_zI_z)$. $[N]=(N_y+1)/2$ is used to locate the origin of coordinate in the center of the ribbon.  $\pi \Phi$, the Peierls phase shift, results from the magnetic flux passing half a hexagonal ring. The diagonal term $\gamma_6\,+\beta^2\gamma_5\,/2$  ($\beta^2\gamma_2\,/2$), due to the interribbon  interactions,  can be treated as the equivalent site energy of $A$ $(B)$ atoms. The centers of the two ribbons in a unit cell  do not coincide with each other [Fig. 1(a)]. This leads to a different tridiagonal matrix $h_2$, which is
\begin{eqnarray}
\begin{cases}
h_2(2m-1,2m-1)=\gamma_6\,+\beta^2\gamma_5\,/2,\\
h_2(2m,2m)=\beta^2\gamma_2\,/2,\\
h_2(2m-1,2m)=h_2(2m,2m-1)=2\gamma_0\,cos\{\sqrt 3\,bk_x/2-\pi( m-[N] +\frac {1}{3}) \Phi\},      \\
h_2(2m-1,2m+2)=h_2(2m+2,2m-1)=\gamma_0.                  \\
\end{cases}
\label{h2}
\end{eqnarray}
The term $\pi\Phi/3 $  in Eq. (\ref{h2}) is caused by the difference between the centers of the two ribbons in a unit cell.

The two off-diagonal block matrices satisfy a simple relation $h_{12}=h_{21}$. As a result of  the interribbon interactions, the nonzero matrix elements $h_{12}(i,j)$ of the off-diagonal block matrix $h_{12}$ are as follows:
\begin{eqnarray}
\begin{cases}
h_{12}(2m-1,2m-1)=\beta\gamma_1,\\
h_{12}(2m,2m)=\beta\gamma_3,    \\
h_{12}(2m-1,2m)=2\beta \gamma_4\,cos\{\sqrt 3\,bk_x/2- \pi( m-[N]+\frac{1}{3}) \Phi\},                  \\
h_{12}(2m,2m-1)=2\beta\gamma_4\,cos\{\sqrt 3\,bk_x/2- \pi( m-[N]) \Phi\},\\
h_{12}(2m,2m+1)=\beta\gamma_4          ,\\
h_{12}(2m-1,2m+2)=\beta\gamma_4, \\
h_{12}(2m,2m+2)=2\beta\gamma_3\,cos\{\sqrt 3\,bk_x/2-\pi( m-[N]+\frac{1}{6}) \Phi\}. \\
\end{cases}
\label{h12}
\end{eqnarray}
The effect of magnetic fields is reflected in the interaction $2\gamma_i cos(\sqrt 3\,bk_x/2 -
\pi(m-[N])\Phi)$. All the nonzero elements of $h_{12}$ depend on $\beta=2 cos(k_zI_z)$. The interribbon interactions are equal to zero at $k_zI_Z=\pi/2$, and they are maximum at $k_zI_z=0$. Eigenvalues and  eigenvectors are obtained after the diagonalization of the Hamiltonian matrix.  Eigenvalues are the  energy dispersions $E^{c,v}(k_xI_x,k_zI_z)$, where $c$ ($v$) represents the unoccupied (occupied) states.  The eigenvector
$|{\emph{a}}^{c,v}_{A^1_1},
  {\emph{b}}^{c,v}_{B^1_1},
  \cdots,
 {\emph{a}}^{c,v}_{A^1_m},
 {\emph{b}}^{c,v}_{B^1_m},
 \cdots,
 {\emph{a}}^{c,v}_{A^2_1},
 {\emph{b}}^{c,v}_{B^2_1},
 \cdots,
 {\emph{a}}^{c,v}_{A^2_m},
 {\emph{b}}^{c,v}_{B^2_m},
 \cdots\rangle$
 is the $k_z$-dependent envelope function $|\Psi^{c,v}(y,k_zI_z)\rangle$ along the $y$ direction.

\section{Magneto-electronic properties}

The band structures of the 2D multilayer graphene ribbons exhibit a highly anisotropic structure. Figs. 2(a)-2(c) show the $k_x$-dependent low-energy bands for the AB-stacked zigzag graphene ribbon with width $N_y$=3000 (639 nm) at
${\rm B}=0$, ${\rm B}=20$ T, and ${\rm B}=10$ T, respectively.
The unit of energy is eV. The Fermi level is set to $E_F=0$. At ${\rm B}=0$, the low energy bands are similar to those obtained from the LDA calculations.\cite{Miyamoto} These bands include parabolic bands and flatbands [Fig. 2(a)]. The occupied valence bands are asymmetric to the unoccupied conduction bands about $E_F=0$ due to the energy-dependent interribbon interactions. The $N_y=3000$ AB-stacked zigzag graphene ribbon is a semimetal because the valence bands lightly  touch the conduction bands at $E_F=0$. The degeneracy of the flatbands at $E=0$, a special feature of a single zigzag ribbon due to the zigzag-edge boundaries, is lifted by the interribbon interactions. The partial flatbands near $E_F=0$ correspond to the edge states mainly localized at the outmost zigzag positions. According to Figs. 2(b)-2(d), the magnetic field drastically modifies the energy bands, such as the alteration of the band feature, the shift of the subbands, and the production of the Landau levels. At ${\rm B}=20$ T, as shown in Fig. 2(b), the original parabolic bands might become the complete Landau levels. The occupied states $E^v$ are asymmetric to the unoccupied states $E^c$ about $E_F=0$. The Landau-level energies do not follow the simple relation $E \propto \sqrt{|n|{\rm B}}$. The presence of $\bf B$ induces a longer and weak splitting energy subband near $E_F$, namely, the zero mode. There are more Landau subbands and a shorter range in band structure as the magnitude of the magnetic field decreases [Fig. 2(c)]. The zigzag boundaries make narrow graphene ribbons exhibit special electronic and magnetic properties.\cite{Lee,Son}

Interestingly, the interribbon interactions induce the concave-upward and -downward magnetobands along $\hat k_z$. The energy dispersions near $E_F$ at ${\rm B}=20$ T with $N_y=3000$, ${\rm B}=10$ T with $N_y=3000$, and ${\rm B}=20$ T with $N_y=6000$ (1278 nm) along the $k_z$-axis are shown in Figs. 2(e)-2(h), respectively, to illustrate the effect of interribbon interactions. There are oscillating subbands, concave-upward and -downward bands,  and nearly dispersionless bands nearby $E_F$ [Figs. 2(e) and 2(f)]. The occupied states $E^v$ are asymmetric to the unoccupied states $E^c$ about $E=0$. The main feature of the energy dispersions remains unchanged for the variation of magnetic flux. However, the state energies are sensitive to the magnitude of the magnetic field [Fig. 2(g)]. The zero mode near $E_F$, including two oscillating subbands, has a dispersion of 0.04 eV at the low field limit. The upper subband of the zero mode, indexed by $n=0$, shows the cosine energy bands $E\propto cos(k_zI_z)$ along the $k_z$ directions, while the feature of the lower subband, indexed by $n'=0$, is distorted. A pseudogap $\Delta$ associated with the zero mode near $E_F$ is observed. Thus, the size of the pseudogap  $\Delta$ varies with $k_z$. The upper limit of the pseudogap  $\Delta$, occurring  at $k_zI_z=\pi/2$, is  equal to $\gamma_6$, the chemical difference  between $A$ atoms and $B$ atoms. The concave-upward and -downward subbands,  are classified into two groups. One group, denoted as Group I, moves towards $E_F$ while the other, Group II, moves away from $E_F$. The first three subbands of each group are shown in Fig. 2(f) as examples. The unoccupied states $E^c$ and occupied states $E^v$ of Group I are denoted by subband indices $n$ and $n'$, respectively. Those of Group II  are indexed, respectively, by $\bar n$ and $\bar n'$. The subband crossings occur at different $k_z$'s. The band edge states, the local minimum and maximum, of the $n$, $\bar n$, and $\bar n'$ subbands are located at $k_{ze}=0$.


The Landau plot, the Landau-level energies vs. the field strength, at different $k_z$'s can reveal the effect of the interribbon interactions. The Landau plot of the $N_y=3000$ AB-stacked zigzag ribbon at $k_zI_z=\pi/2$ is shown in Figs. 3(a) and 3(b). The state energies of the zero mode are independent of the field strength. The energy spacing between the upper and lower subbands, the pseudogap $\Delta$ at $k_zI_z=\pi/2$, is equal to $\gamma_6$.  Except the zero mode, Landau-level energies  follow the simple relation $E^{c,v} \propto \sqrt{\rm B}$.  The chief reason  is that the two ribbons in a primitive unit cell are regarded as two independent ribbons because the interribbon  interactions  are turned off at $k_zI_z=\pi/2$.  The magnetoelectronic properties are dominated by the magnetic confinement rather than the quantum effect. Accordingly, the Landau-level
energies, indexed by $n$, follow the relation
$E \propto{\sqrt {|n|\rm B}}$.\cite{Huang02} On the other hand, the interribbon interactions  between the two ribbons are maximum at $k_zI_z=0$. The Landau plots, as shown in Figs. 3(c) and 3(d),  exhibit  different behaviors. The state energies of the zero mode vary with the field strength. In the energy region $|E| < 0.02\gamma_0 \simeq 50$ meV, the state energies of the first and second unoccupied (occupied) Landau levels  exhibit the linear-in-B dependence at $ B \le 50~ {\rm T}$. The high energy Landau levels, indexed by  $n \geq 3$, exhibit the linear-in-B dependence at $ B < 15~ {\rm T}$. Then, they evolve from a linear B-dependence to a square-root B dependence with the increase in the field strength. Finally  they  exhibit the linear-in-$\sqrt {\rm B}$ dependence at ${\rm \bf B} \ge 20 {\rm T}$.

Density of state (DOS) of multilayer graphene ribbons is evaluated by ${D(\omega)= \sum_{{\bf k},h}}$ ${| \nabla_{\bf k}E^h({\bf k};\phi) |^{-1}_{E^h=\omega }}$. As shown in Fig. 4(a), in the absence of a magnetic field, the featureless
DOS of the $N_y=3000$ graphene ribbon, asymmetric about $\omega=0$, shows a small bump at $\omega =0$, which results from the tiny overlap between the occupied bands and unoccupied bands [Fig. 2(a)]. The $N_y=3000$ graphene ribbon is a semimetal. The magnetic field modifies electronic states [Fig. 2] and the aspect of DOS  [Fig. 4(b)].  DOS at ${\rm B}=20$ T is asymmetric about $E_F$. It chiefly exhibits three kinds of structures, including the delta-function-like peak, the compound peak near $\omega=0$, and the 1D power-law divergences. The first kind of peaks
are related to the Landau levels
along $\hat k_x$ [Fig. 2(b)] and dispersionless bands along
$\hat k_z$ [the red curves in Fig. 4(b)], the second kind
to compound band along $\hat k_x$ and $\hat k_z$, and
the third kind to concave-upward and -downward bands
along  $\hat k_z$ [the red curves in Fig. 4(b)].
The sharp peaks at $\omega =-0.3$ and $-0.45$ eV origin in the Landau levels along $\hat k_x$ and dispersionless bands along $\hat k_z$. DOS of the
concave-upward and -downward diverges in the forms ${1/\sqrt {\omega\,-E(k_{ze})}}$ and ${1/\sqrt {E(k_{ze})-\omega}}$, respectively, where $k_{ze}$ is the band edge of the subband. DOS exhibits more peaks in the decreases of the magnitude of magnetic fields, as shown in Fig. 4(c). In other words, the peak heights and positions are strongly dependent on the magnetic flux. Fig. 4(d) shows DOS of the $N_y=6000$ graphene ribbon at ${\rm B}=20$ T. Remarkably, density of states is independent of the ribbon width [Fig. 4(b) and Fig. 4(d)].

When the ribbon width is sufficiently large ($N_y\ge 3000$), the AB-stacked graphene ribbons at ${\rm B=0}$ are expected to be almost identical to the bulk graphite in DOS except at very low frequency. The zigzag boundary induces  partial flat bands near $E_F=0$ and thus strongly affects the low-frequency physical properties. The two kinds of systems might differ from each other in the transport and magnetic properties. Nevertheless, the large graphene ribbons are useful in understanding the bulk graphite, e.g., magnetoelectronic properties\cite{Kopelevich} and magneto-optical absorption spectra. The main reason is that the calculations of the magnetoelectronic structures are very complicated for the latter at ${\rm B}$ $<$ 10 T (a very large Hamiltonian matrix). That is to say,  the magnetoelectronic properties (state energies and  associated wave functions) remain the same when the ribbon width is sufficiently wide. It is deduced to be the same with that of a 3D bulk graphite.

The effect of interribbon interactions on the envelope functions deserves a closer examination. Envelope functions of the $N_y=3000$ zigzag ribbon subjected to ${\rm B=20}$ T at different $k_z$'s are shown in Figs. 5(a)-5(d). Each envelope function $\Psi(y,k_zI_z)$ is separated into eight components, $ \psi(A^1_o)$, $\psi(B^1_o)$, $\psi(A^1_e)$, $\psi(B^1_e)$, $\psi(A^2_o)$, $\psi(B^2_o)$, $\psi(A^2_e)$, and $\psi(B^2_e)$. $A^1_o$ ($B^2_e$) denotes $A$ ($B$) atoms located at the odd (even) zigzag lines at the lower (upper) ribbon plane.  The subenvelope function $\psi(A^1_o)$, for example, is $| {\emph{a}}_{A^1_1},  { \emph{a}}_{A^1_3},{\emph{a}}_{A^1_5},\cdots \rangle$. Because $\psi(A^1_o)=-\psi(A^1_e)$, $\psi(B^1_o)=-\psi(B^1_e)$, $\psi(A^2_o)=-\psi(A^2_e)$, and $\psi(B^2_o)=-\psi(B^2_e)$, only four components  $\psi(A^1_o)$, $\psi(B^1_o)$,  $\psi(A^2_o)$, and $\psi(B^2_o)$ are displayed. At $k_zI_z=\pi/2$, the interribbon interactions are closed due to $\beta= 2cos(k_zI_z)=0$ [Eqs. (\ref{h1})-(\ref{h12})].  The two ribbons in a primitive unit cell are decoupled, i.e., there are two isolated graphene ribbons. Except the zero mode, the state energies are double degenerate. Their associated envelope functions, $\Psi_{1}(y,k_zI_z=\pi/2)$ and $\Psi_{2}(y,k_zI_z=\pi/2)$, are orthogonal to each other and different in the sign of some components.
One of the two envelope functions is shown in Fig. 5(a).
It is exactly described as
$\Big[\phi_{n}(A^1_o),
       \phi_{n-1} (B^1_o),
       \phi_{n-1} (A^2_o),
       \phi_{n} (B^2_o)
\Big],$
where $\phi_{n}$ is the harmonic oscillator, product of the Hermite polynomial $H_{n}$ and the Gaussian function.\cite{Huang02}  Each $\phi_{n}$ has $n$ nodes, where the subenvelope function changes the signs. The node number  $n$ of the subenvelope function
$\psi_{n}(A^1_o)$ is identical to the subband index $n$. Thus, we use the subband indices $n=1,~2,\cdots$ and $n'=1,~2,\cdots$ to specify the characteristic of envelope function [Fig. 5(a)]. $\phi_{n} (A^1_o)$ and $\phi_{n}(B^2_o)$ [$\phi_{ n-1} (B^1_o)$ and $\phi_{ n-1} (A^2_e)$] belong to the $n$  [$n-1$] mode. The envelope functions are the same as those of a single graphene ribbon.\cite{Huang02} Due to the orthogonality,  $\Psi_{n}$
and $\Psi_{n'}$ are different in the sign of some  components.
Besides, $\Psi_{n=0}=\phi_{0}(B^2_o)-\phi_{0}(B^2_e)$,
the envelop function of the $n=0$ subband, is located in the $B$ sites of the upper ribbon. $\phi_{0}(A^1_o)-\phi_{0}(A^1_e)$, the envelop function of the $n'=0$ subband, is mainly controlled by the $A$ atoms on the lower ribbon plane. The chemical difference $\gamma_6$ changes the site energies of atoms $A$ and, thus, causes the energy split and produces a pseudogap  $\Delta=\gamma_6$ between  $\Psi_{n=0}$ and $\Psi_{n'=0}$ at $k_zI_z=2/\pi$ [Fig. 2(e)].

Away from  $k_zI_z=\pi/2$, the switched-on interribbon interactions not only produce  two groups of subbands, but also change the aspects of envelop functions. At
$k_zI_z=0.499\pi$, the envelope functions related to $n$
and $\bar n$ subbands concentrate on the lower  and upper
ribbon, respectively  [Fig. 5(b)].  The subenvelope
functions related to the $n$ subband are $\phi_{n}(A^1_o)$
and $\phi_{n-1} (B^1_o)$, where $n=1,~2,~3, \cdots$
 ($n'= 1,~2,~3, \cdots$). As for the $\bar n$  subbands, the associated subenvelope functions are $\phi_{\bar n} (A^2_o)$ and  $\phi_{\bar n+1} (B^2_o)$, where
$\bar n =0,~ 1,~2,~3\cdots$. Notably, $\bar n$ is different from $n$ and it begins from  $\bar n =0$ ($\bar n' =0$). For example, the envelop function of $n=1$ [$\bar n=0$] subband is $\Psi_{n=1}=\phi_{1}(A^1_o)+\phi_{0}(B^1_o)$
[$\Psi_{\bar n=0}=\phi_{0}(A^2_o)+\phi_{1}(B^2_o)$].
The effect caused by the interribbon interactions on the envelope functions can be interpreted by the first order perturbation theory. Around  $k_zI_z=0.5\pi$, the matrix elements of $h_{12}$ are proportional to  $(\pi/2 -k_zI_z)$ because of  $\beta=2 cos(k_zI_z)\sim 2(\pi/2 -k_zI_z)$. Thus, the weak interribbon interactions lift the degeneracy at $k_zI_z=\pi/2$ and produces a two-level system, for example,
$n=1$ and $\bar n=0$ subbands. The related  envelope functions are $\Psi(y,k_zI_z)=[\Psi_{1}(y,k_zI_z=\pi/2)\pm
\Psi_{2}(y,k_zI_z=\pi/2)]/\sqrt{2}$, the bonding or antibonding of  the envelope functions at $k_zI_z=\pi/2$.
As a result, the envelope functions of the $n$ ($\bar n$)
subbands chiefly distribute on the lower (upper) ribbon with the different spatial symmetries. In addition, interribbon interactions have no influence on the envelop functions  $\Psi_{n=0}$ and $\Psi_{n'=0}$.

The  feature of the wave function at $k_zI_z=0.4\pi$, as shown in Fig. 5(c), is dissimilar to that at $k_zI_z=0.499\pi$ [Fig. 5(b)].  The charge carriers of the former distribute on two ribbon planes and each subenvelope function is of the same importance, whereas those of the latter concentrate only on one ribbon plane. The varying interribbon interactions causes the redistribution of charge density. The definition of subband
indices $n$ and $\bar n$ ($n'$ and $\bar n'$) is based on
the spatial symmetry of the subenvelope function
$\psi_{n}(A^1_o)$. The subenvelope functions are
proportional to the harmonic oscillator $\phi_{n}$,
for example,  $\psi_{n=1}(A^1_o)\sim  \phi_{ 1}(A^1_o) $.
The envelope functions of the $|n|\geq 1$ subband are
$\big[\phi_{n}(A^1_o),\phi_{n-1}(B^1_o),\phi_{n}(A^2_o), \phi_{n+1}(B^2_o)\big]$. The subband index $n$ ($n'$) is assigned to be  $n=1,~2,~3, \cdots$  ($n'= 1,~2,~3, \cdots$). Notably, subenvelope functions $\psi(A^1_o)$ and $\psi(A^2_o)$ belong to the $n$ mode and $\psi(B^1_o)$ [$\phi (B^2_o)$] to the $n-1$ [$n+1$] mode. The spatial symmetry $[n, n-1, n, n+1]$ is different from
that $[n,n-1, n-1, n]$  at $k_z=0$. On the other hand,
the envelope functions of the
$\bar n$ subband is the linear combination of  components
$\phi_{\bar n}(A^1_o)$, $\phi_{\bar n-1}(B^1_o)$,
$\phi_{\bar n}(A^2_o)$ , and $\phi_{\bar n+1}(B^2_o)$.
The  spatial symmetry $[{\bar n}, {\bar n-1} ,{\bar n},
{\bar n+1}]$ of the $\bar n$ subband is the same as that
of the $n$ subband $[n, n-1, n, n+1]$. According to the
spatial symmetry of $\psi_{n}(A^1_o)$, the subband indices
are assigned as $\bar n=0,~1,~2,\cdots$. Notably, $\bar n$
starts at zero. Due to the interactions between the $n=\pm 6$ and $\bar n=\pm 1$ subbands, the shapes of $\bar n=0$
envelope functions is modified and its component
$\psi(B^1_o)$ almost disappears.

From  $k_zI_z=0.4\pi$ to $k_zI_z=0$, the gradually enhancing interribbon interactions significantly modify  the feature or the site amplitude of the $n$ envelope functions, as shown in Fig. 5(d). The maximum interribbon
interactions, occurring at $k_zI_z=0$, dramatically change the main feature of the envelope function [heavy dots
in Fig. 5(d)], which now can not simply be described by
harmonic oscillator $\phi_{n}$. For comparison, those at
$k_zI_z=0.3\pi$ are shown by light dots in Fig. 5(d), whose main feature of the envelope functions is still similar to those at $k_zI_z=0.4\pi$. Thus, the envelope function is approximately linear superposition of components $\phi_{n}(A^1_o)$, $\phi_{n-1}(B^1_o)$, $\phi_{n}(A^2_o)$,
and $\phi_{n+1}(B^2_o)$. The interribbon interactions dramatically alter the site amplitude.  $\psi^{c,v}_{n-1}(B^1_o)$  and  $\psi^{c,v}_{n+1}(B^2_o)$ dominate the envelope function. The comparison shows that
the increase in the interribbon interactions not only
change the main feature of the subenvelope functions
but also enhance the site amplitude of $B$ atoms. It should
be noted that the interribbon interactions do not break
the spatial symmetry of the envelope function, which is
denoted as $\big[\psi_{n}(A^1_o), \psi_{n-1}(B^1_o),  \psi_{n}(A^2_o), \psi_{n+1}(B^2_o)\big]$.
The spatial symmetry of the envelope function has a critical effect on the optical  selection rule.

The  dependence of magnetoelectronic properties of the  $\bar n$ subbands on the wave vector $k_z$ is deliberated. For comparison purpose, the  $k_z$-dependent  state energies of the $n=1$ and $\bar n=0$ subbands and the associated envelop functions  are shown in Fig. 6.  At $k_z=\pi/2$, two subbands  are degenerate and magnetoelectronic properties mainly exhibit the characteristics of an isolated graphene because of the close of the interribbon interactions. As $k_z$ changes from $\pi/2$ to zero, the two degenerate subbands are lifted by the $k_z$-dependent interribbon interactions and form concave-upward and -downward  bands along  $k_z$ [Fig. 6(b)]. Meanwhile, the interribbon interactions have different influences  on the shape of wave functions  and  the charge-carrier distribution [Fig. 6(a)]. The subenvelope functions of $\bar n=0$ subband [heavy dots in Fig. 6(a)] can be clearly described by the harmonic oscillator $\phi_n$. By contrast, the features of subenvelope functions related to $n=1$ subband are seriously modified
by the $k_z$-dependent interribbon interactions as
$k_zI_z \rightarrow 0$. The $k_z$-dependent interribbon
interactions cause the change of the site amplitudes.  The
charge carriers of the $\bar n=0$ subbands tend to concentrate on $A$ atoms, while those of $n=1$ subbands are mainly located in $B$ atoms  [Fig. 6(a)].

The characteristics of wave functions are applicable to  determine the optical transition channels.  The optical absorption spectra is $A(\omega)\propto
\sum_{\mu,\nu} \int_{1stBZ}\frac{dk_{x}}{2\pi} ~\frac{dk_{z}}{2\pi}D(\omega)\times |M_{\mu,\nu}|^2$, where $\mu=n,~\bar n$ and $\nu=n',~\bar n'$ are subband indices. $D(\omega)$ is the joint density of states and $M_{\mu,\nu}$ is the velocity matrix element.\cite{Huang01,Huang02,Huang03}  According to the result of Ref. \cite{Huang03}, the magnitude of velocity matrix is $|M_{\mu,\nu}|= |M_{\mu,\nu}|_{intraribbon} + |M_{\mu,\nu}|_{inter-ribbon}$, where $|M_{\mu,\nu}|$ is proportional to $\langle \Psi_{\mu}(y,k_zI_z) |\Psi_{\nu}(y,k_zI_z)\rangle$, the projection of the envelope function of initial state,  $|\Psi_{\nu}(y,k_zI_z)\rangle$, onto that of final state, $|\Psi_{\mu}(y,k_zI_z)\rangle$. The subscript $intraribbon$ ({\it inter-ribbon}) denotes the initial and final states located at the same (different) ribbons. The $intraribbon$-projection, for example, is \begin{eqnarray}
\langle \Psi_{n}(y,k_zI_z) |\Psi_{n'}(y,k_zI_z)\rangle_{intraribbon}
 &=& \langle \psi_n(A^1)|\psi_{n'}(B^1)\rangle +\langle \psi_{n}(B^1)| \psi_{n'}(A^1)\rangle +\\
 & & \langle \psi_n(A^2)|\psi_{n'}(B^2)\rangle +\langle \psi_{n}(B^2)| \psi_{n'}(A^2)\rangle .
\end{eqnarray}
The velocity matrix element  $M_{\mu,\nu}$ depends on the spatial symmetries of the initial and final states and, thus,  determines the selection rule and the  effective  transition channels.\cite{Huang01,Huang02,Pereira} At $k_zI_z=\pi/2$, only the intraribbon optical transition, $|M_{\mu,\nu}|_{intraribbon}$, makes contribution to the velocity matrix element $|M_{\mu,\nu}|$. There is only one Group of subbands, i.e., $\mu=n$ and $\nu=n'$.  Therefore, the optical selection rule is predicted to be $\delta_{n-n'}=\pm 1$, where $n=0,~1,~2,~3\cdots$ and $n'=0,~1,~2,~3\cdots$. The selection rule  is the same as that of a monolayer graphene ribbon.\cite{Huang02}

Away from $k_zI_z=\pi/2$,  there are four possible optical transition channels $n'\rightarrow n$,  $\bar n'\rightarrow \bar n$, $\bar n'\rightarrow n$ and $n'\rightarrow \bar n$ [Fig. 6(b)]. According to the spatial symmetry of  envelope functions, the corresponding selection rules of the four transitions are predicted to be $\delta_{n-n'}=\pm 1$, $\delta_{\bar n-\bar n'}=\pm 1$, $\delta_{ n-\bar n'}=\pm 1$, and $\delta_{\bar n-n'}=\pm 1$. The change of the state energies caused by interribbon interactions might alter the peak positions of joint density of states, and the modification of the shape and site amplitude of envelope functions might give rise to the vivid variation of the magnitude of velocity matrix element $M_{\mu,\nu}$.  Thus, the optical spectra of the AB-stacked ribbon are expected to be more complicated in peak number, peak position and peak height than those of a monolayer or bilayer ribbon. The work to calculate the optical spectra of the AB-stacked ribbon is on-doing.

\section {Conclusions}

The Peierls coupling tight-binding method is employed to investigate the anisotropic magnetoelectronic structures of multilayer graphene ribbons. They are strongly dependent on the $\bf B$-field strength and the interribbon interactions. $\bf B$ can induce Landau levels along $k_x$ and change energy spacing. The interribbon interactions significantly affect state degeneracy, energy dispersions, Landau plot. They destroy the symmetry of magnetoband structures. They also produce  concave and convex subbands along $k_z$. The magnetoband structures are strongly anisotropic. There are many doubly degenerate Landau levels and singlet 1D curving bands along $k_x$ and $k_z$ directions. Such features are directly reflected in density of states. DOS exhibits a lot of asymmetric prominent peaks, for example, the delta-function-like peak, the 1D power-law divergence, and the compound peak. Moreover, the $k_z$-dependent interribbon interactions modify the shape of wave functions, alter the spatial symmetry, and  arouse the change of charge-carrier distribution. The findings are further used to predict the optical selection rule. Most importantly, the large AB-stacked graphene ribbons are expected to be almost identical to the bulk graphite in low-energy  magnetoelectronic properties. Our work  is useful in understanding the bulk graphite, e.g., magnetoelectronic excitations and magneto-optical absorption spectra. The predicted magnetoelectronic properties could be examined by transport and optical measurements.

\begin{acknowledgments}
This work was supported by the Taiwan National Science Council (NSC 96-2112-M-165-001 MY3; NSC 95-2112-M-006-0002).
\end{acknowledgments}

\newpage
\section*{Figure Captions}
\begin{itemize}
\item[FIG. 1.] (a) The geometric structure of the $N_y$=3 AB-stacked zigzag graphene nanoribbons. (b)  $\gamma_0=2.598 eV$ is the intralayer interaction. $\gamma_1=0.364 eV$ ($\gamma_5=0.036 eV$) represents the interaction between two $A$ atoms from two neighboring ribbons (two  next-neighboring ribbons), and $\gamma_3=0.319 eV$ ($\gamma_2=$-0.014 eV) is for $B$ atoms. $\gamma_4= 0.177 eV$ corresponds to the interribbon interaction between $A$ atoms and $B$ atoms.  $\gamma_6=-0.026 eV$ is the chemical-shift between $A$ atoms and $B$ atoms. The values of $\gamma_i$'s are the same as those of AB-stacked graphite.\cite{Charlier}

\item[FIG. 2.] The $k_x$-dependent low-energy bands for the AB-stacked zigzag graphene ribbons with (a) $N_y$=3000 at ${\rm B}=0$, (b) $N_y$=3000 at ${\rm B}=20$ T, (c) $N_y$=3000 at ${\rm B}=10$ T, and (d) $N_y$=6000 at ${\rm B}=20$ T. Figs. (e)-(h), same plot
    as Figs.(a)-(d), but for the $k_z$-dependence.

\item[FIG. 3.]  (a) Landau plot of the $N_y=3000$ ribbon at $k_zI_z=\pi/2$. (b) The Landau level energies of the  $N_y=3000$ ribbon at $k_zI_z=\pi/2$ vs. $\sqrt {\rm B}$. (c) Landau plot of the $N_y=3000$ ribbon at $k_z=0$. (d) The relation between Landau level energies of the $N_y=3000$ ribbon at $k_z=0$ and $\sqrt {\rm B}$.

\item[FIG. 4.]   The low-energy DOS for the AB-stacked zigzag graphene ribbons with (a) $N_y$=3000 at ${\rm B}=0$, (b) $N_y$=3000 at ${\rm B}=20$ T, (c) $N_y$=3000 at ${\rm B}=10$ T, and (d) $N_y$=6000 at ${\rm B}=20$ T. %

\item[FIG. 5.] The envelop functions  for the AB-stacked zigzag graphene ribbons $N_y$=3000 subjected to ${\rm B}=20$ T at $k_zI_z=\pi/2,~ 0.499\pi, ~0.4\pi$ and $0$ are shown, respectively, in (a)-(d).

\item[FIG. 6.] (a)  The energy dispersions of the $n=1$ and $\bar n=0$ subbands. (b) The variation of the
     envelop functions, related to the $n=1$ and $\bar n=0$ subbands, with $k_zI_Z$.
\end{itemize}
\pagestyle{empty}
\newpage
\begin{figure}
\begin{center}
\includegraphics [height=width=1.0\textwidth, height=1.0\textheight]{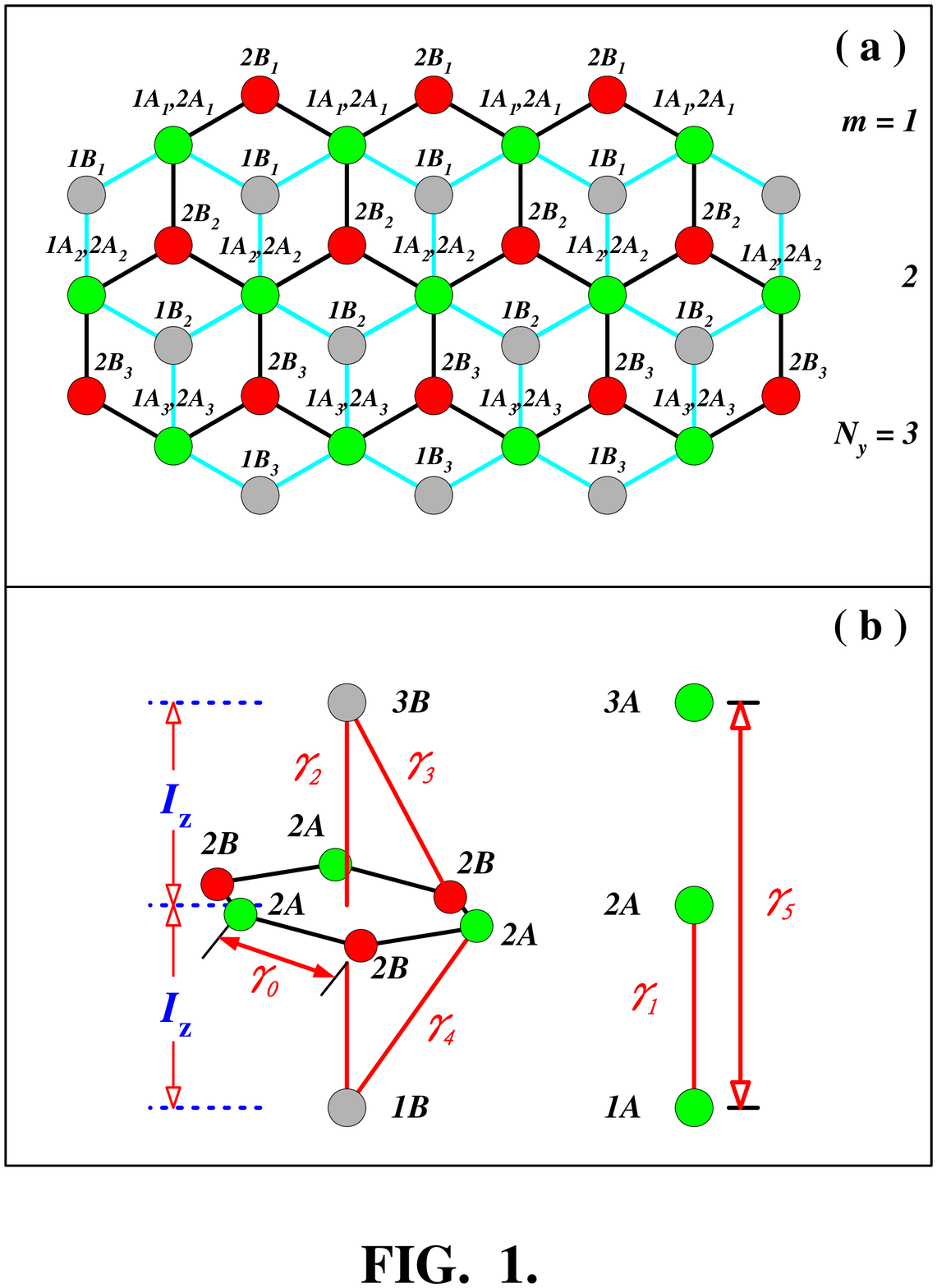}
\end{center}
\end{figure}
\newpage
\begin{figure}[t]
\begin{center}
\rotatebox{180}
{\includegraphics[width=1.0\textwidth, height=1.0\textheight]{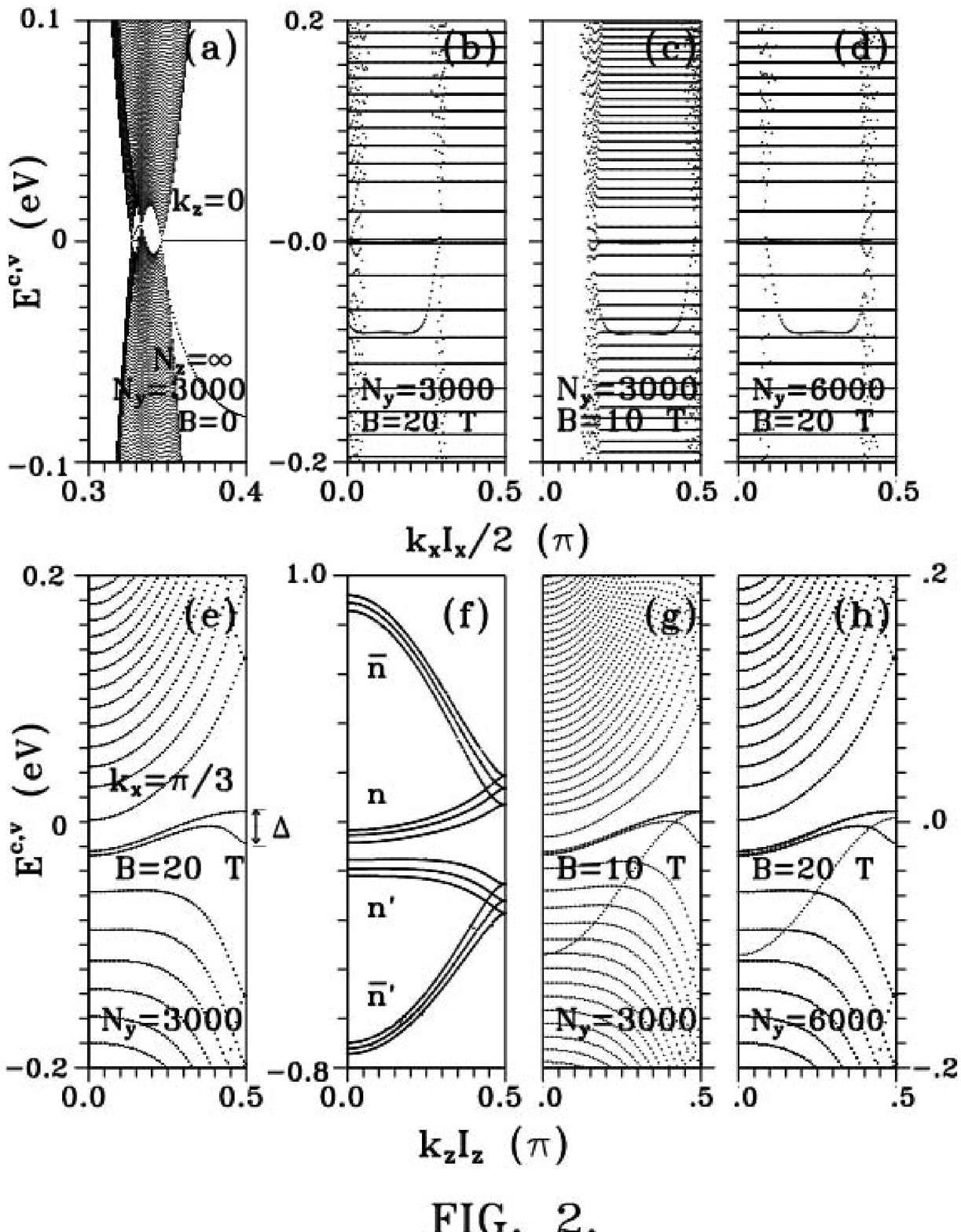}
}
\end{center}
\end{figure}
\newpage
\begin{figure}[t]
\begin{center}
\rotatebox{180}
{\includegraphics[width=1.0\textwidth, height=1.0\textheight]{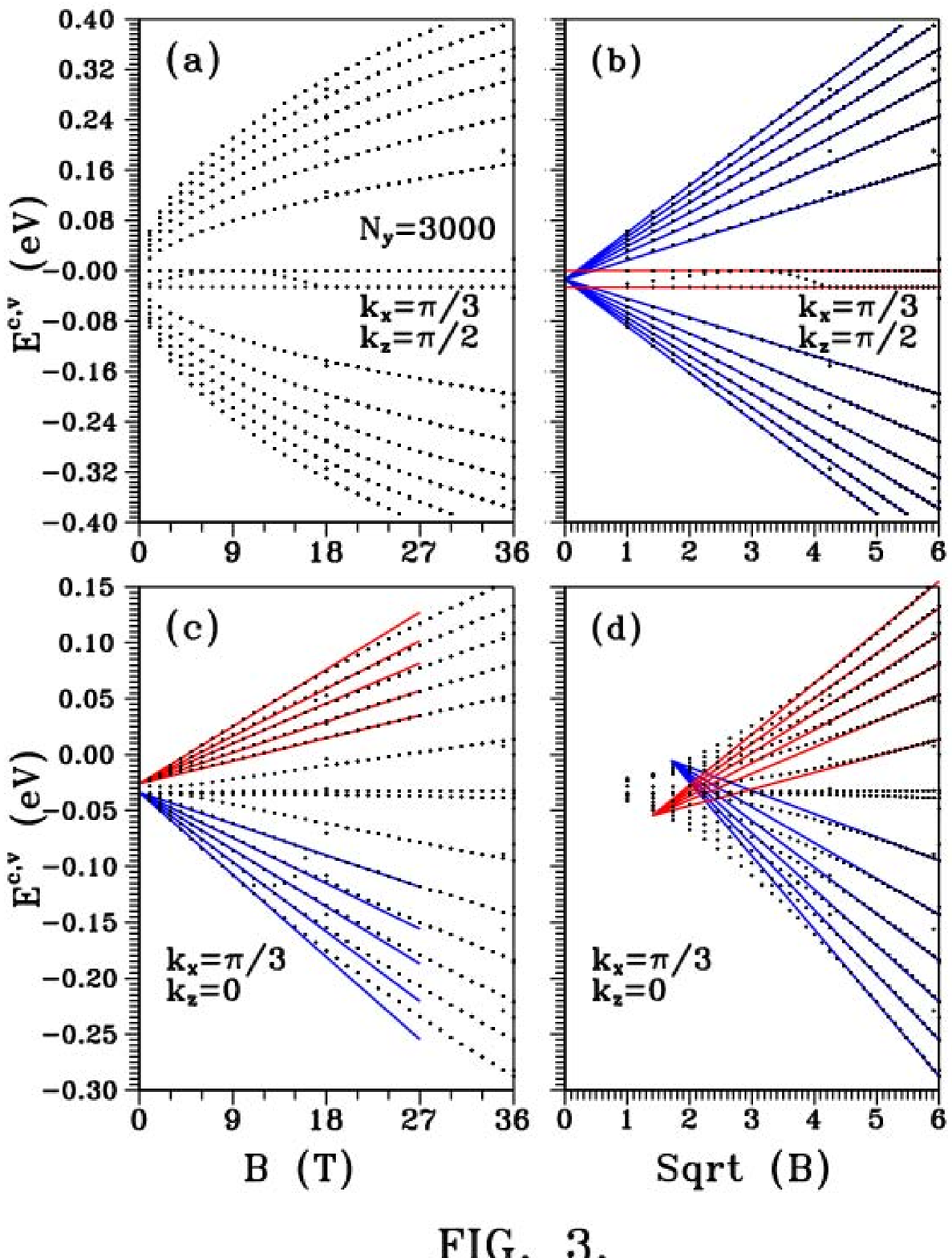}
}
\end{center}
\end{figure}
\newpage
\begin{figure}[t]
\begin{center}3
\rotatebox{180}
{\includegraphics[width=1.0\textwidth, height=1.0\textheight]{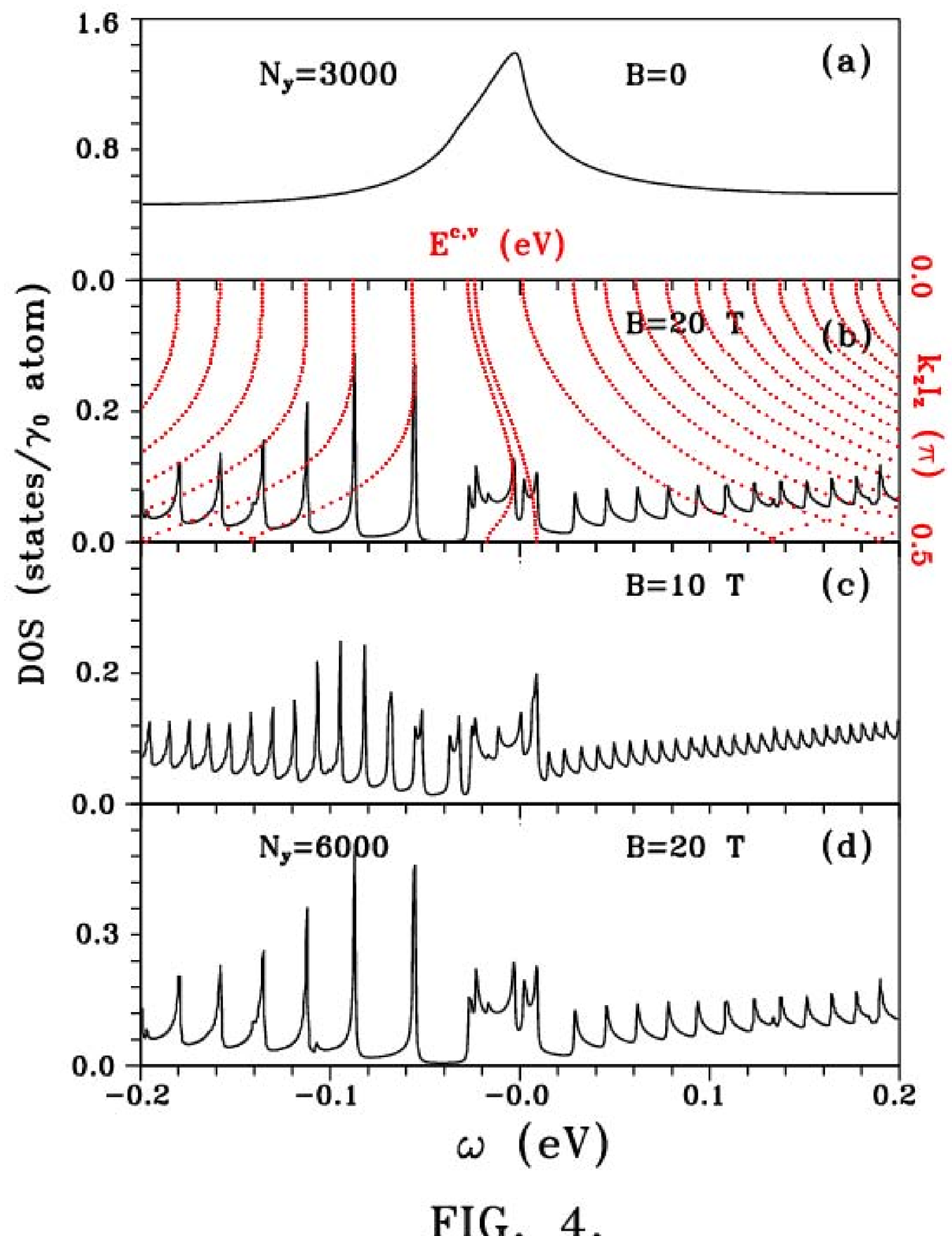}
}
\end{center}
\end{figure}
\newpage
\begin{figure}[t]
\begin{center}
\rotatebox{180}
{\includegraphics[width=1.0\textwidth, height=1.0\textheight]{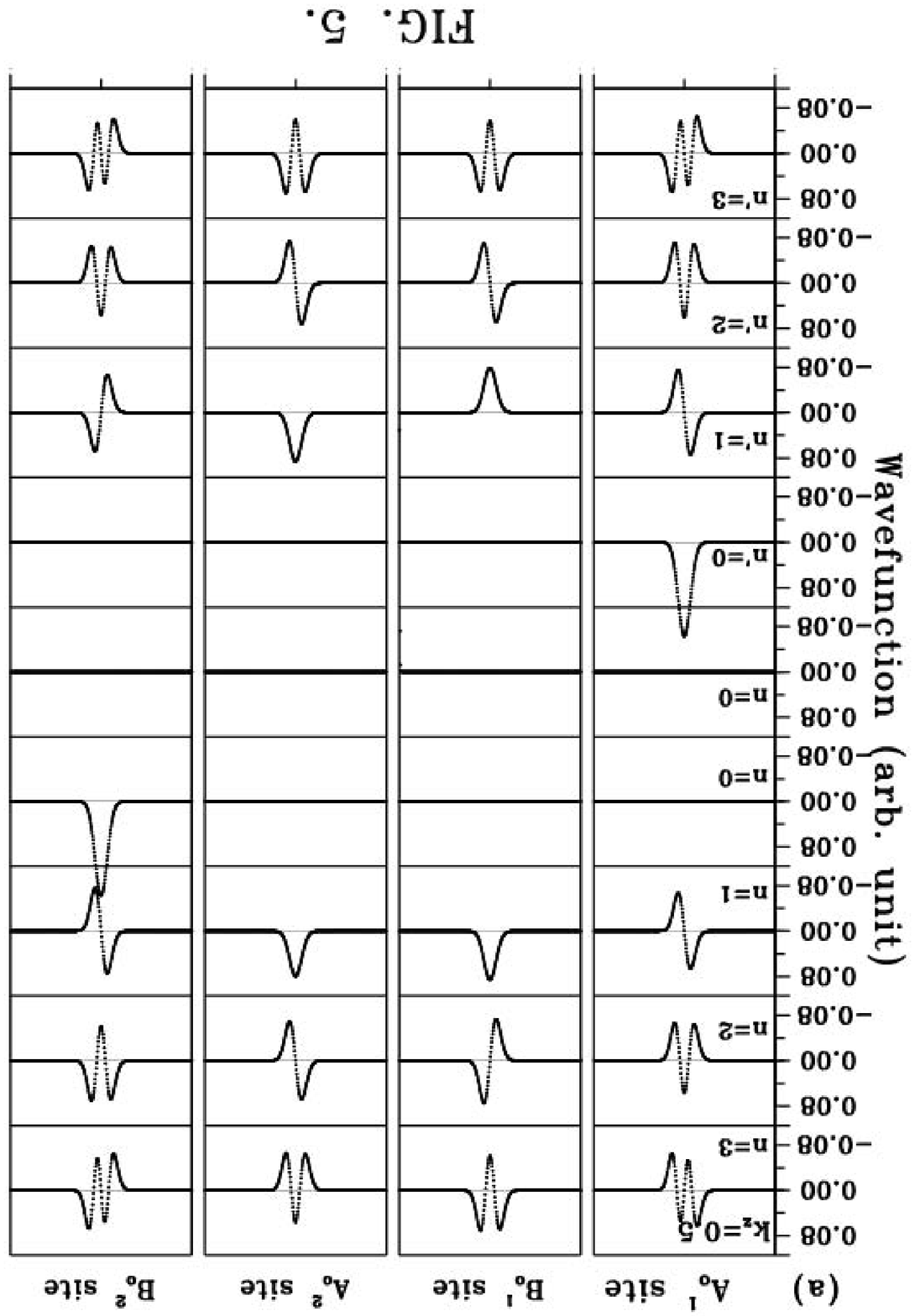}
}
\end{center}
\end{figure}
\newpage
\begin{figure}[t]
\begin{center}
\rotatebox{180}
{\includegraphics[width=1.0\textwidth, height=1.0\textheight]{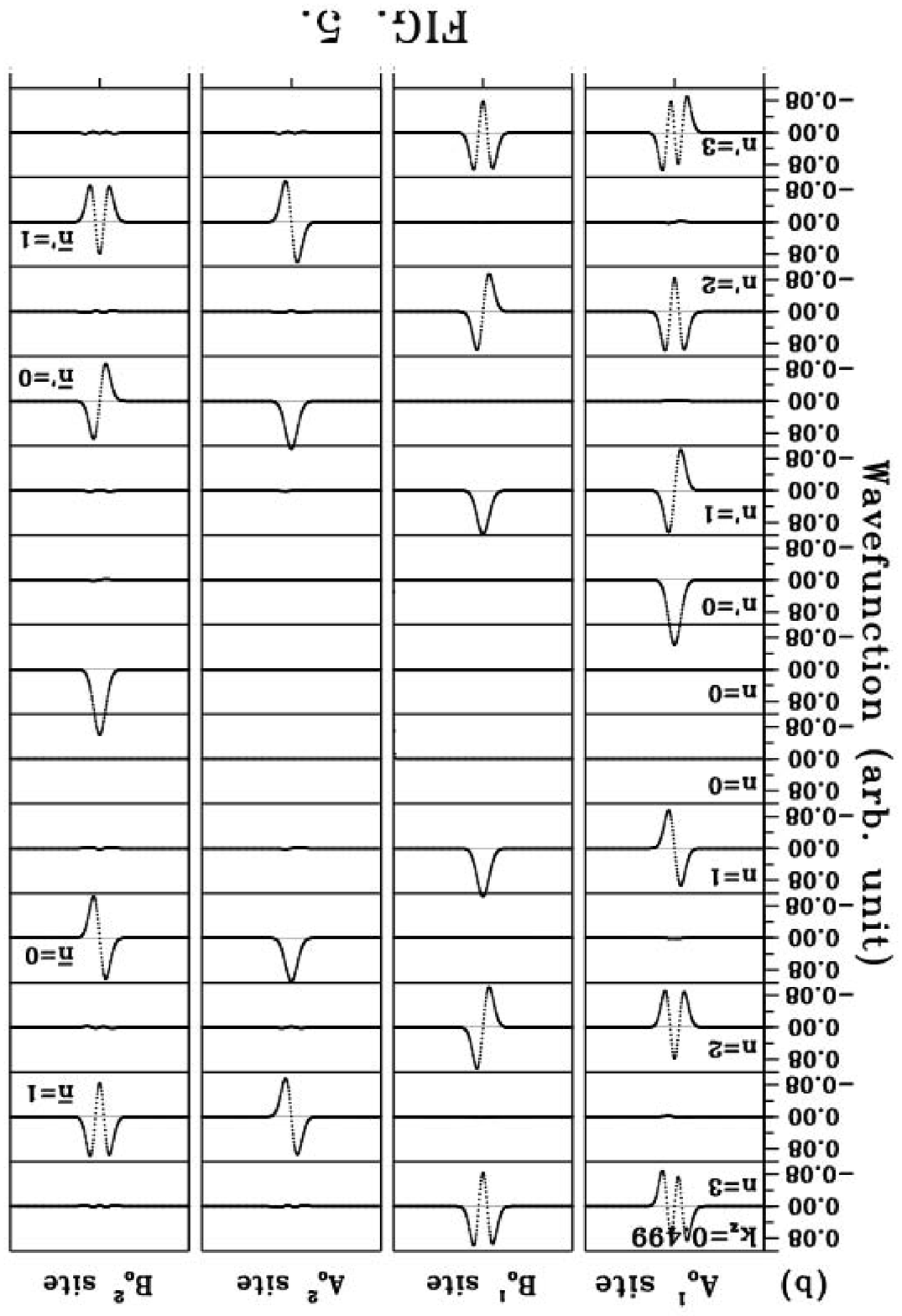}
}
\end{center}
\end{figure}
\newpage
\begin{figure}[t]
\begin{center}
\rotatebox{180}
{\includegraphics[width=1.0\textwidth, height=1.0\textheight]{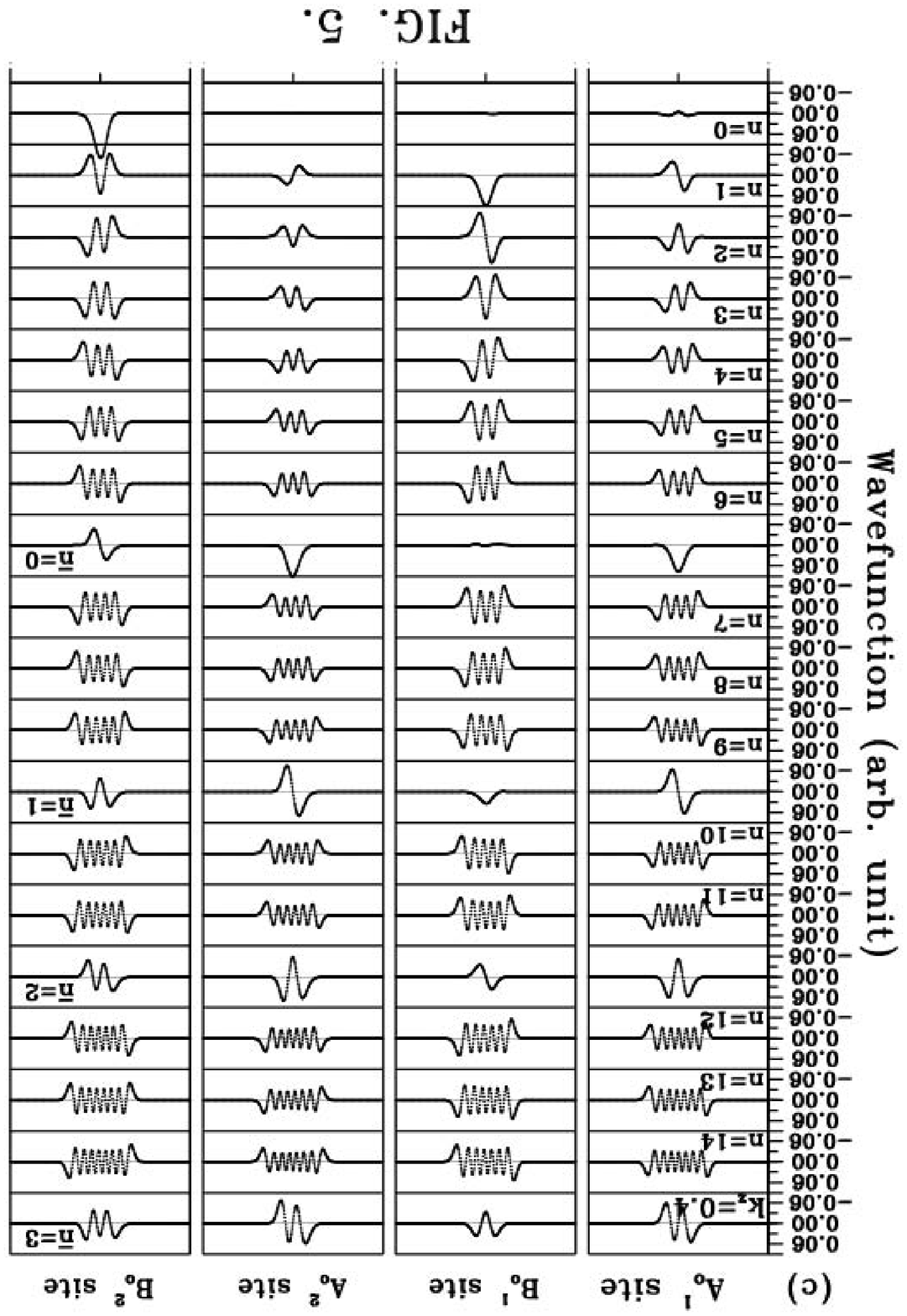}
}
\end{center}
\end{figure}
\newpage
\begin{figure}[t]
\begin{center}
\rotatebox{180}
{\includegraphics[width=1.0\textwidth, height=1.0\textheight]{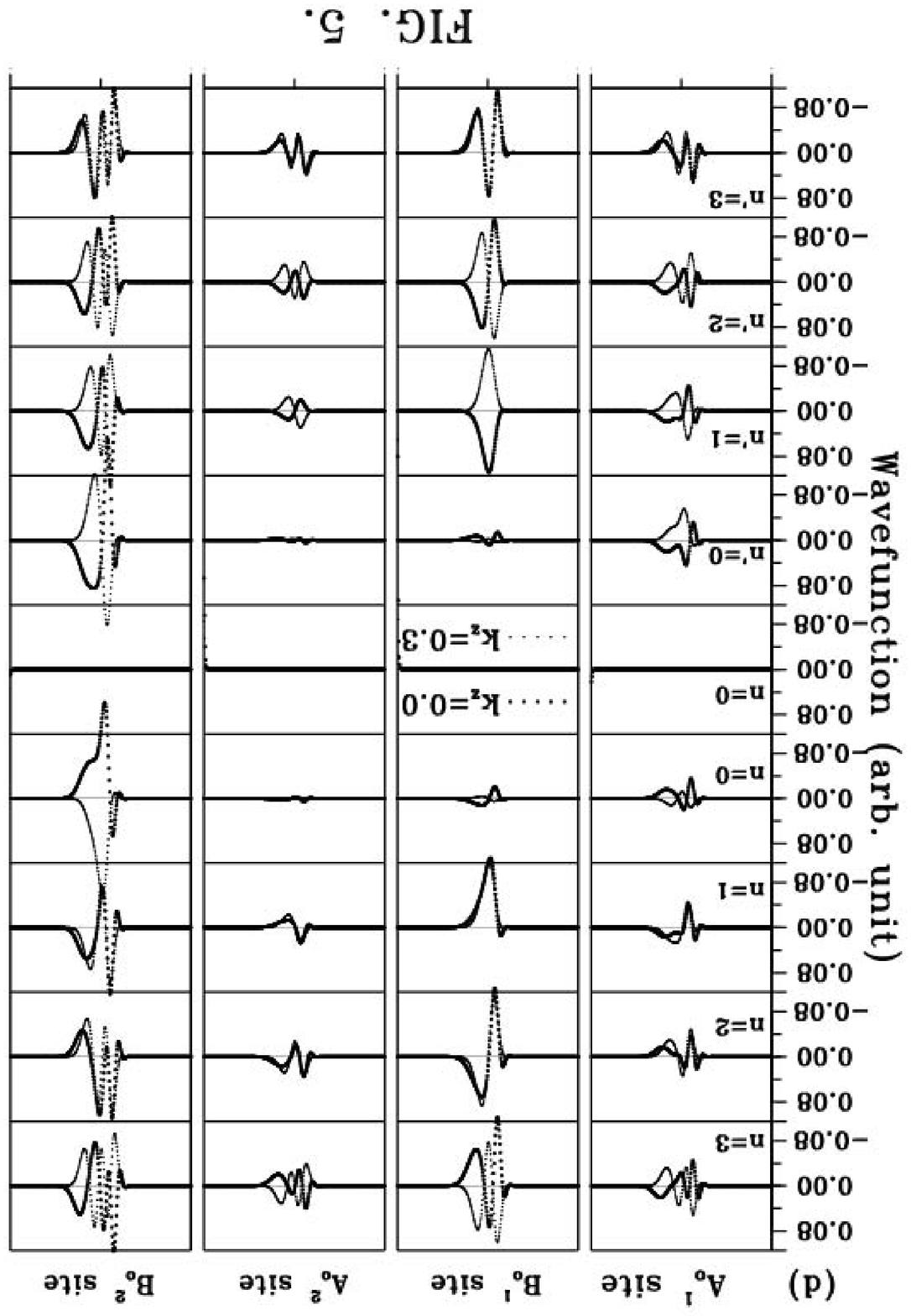}
}
\end{center}
\end{figure}
\newpage
\begin{figure}[t]
\begin{center}
\rotatebox{180}
{\includegraphics [width=1.0\textwidth, height=1.0\textheight]{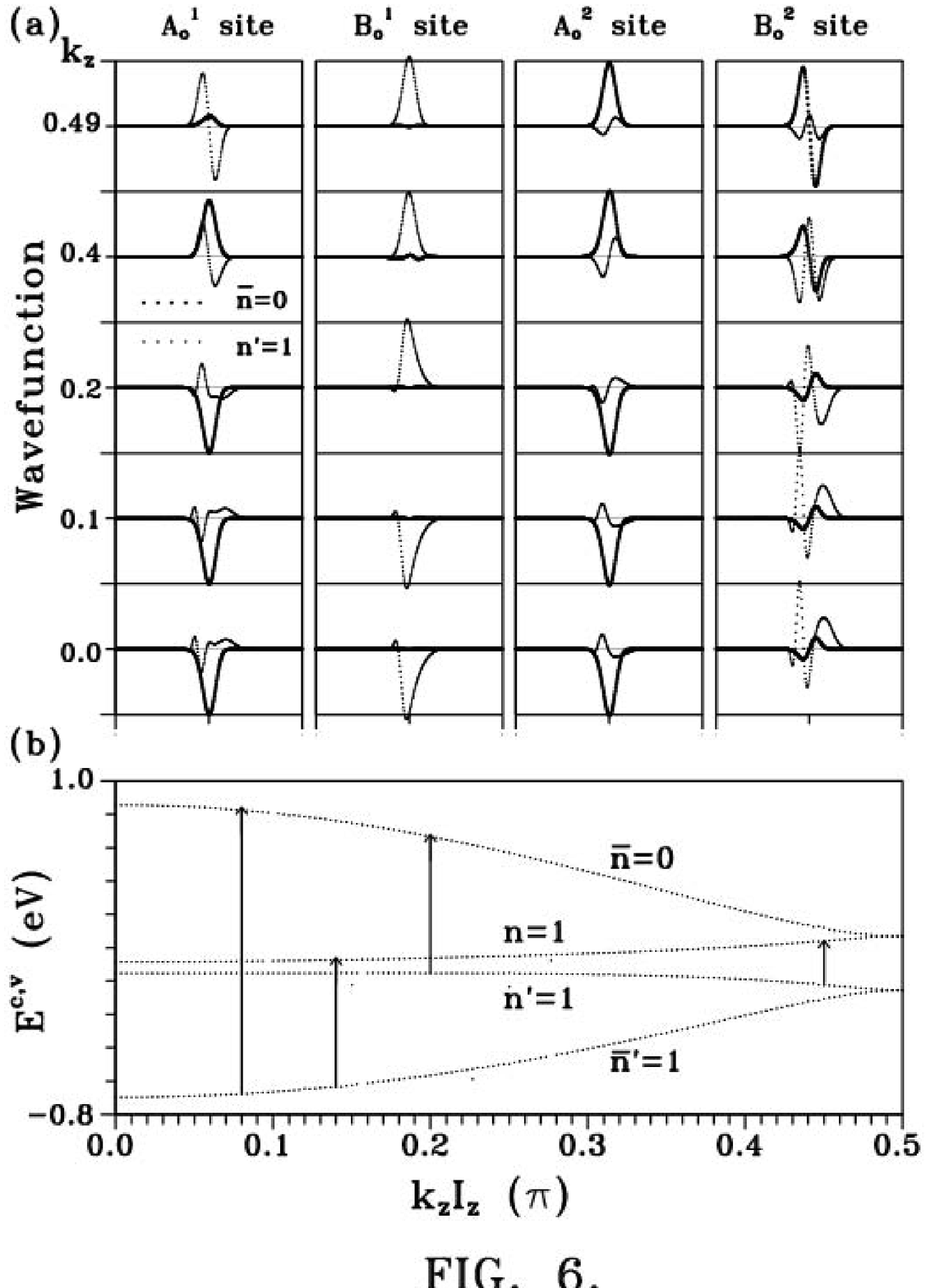}
}
\end{center}
\end{figure}
\end{document}